\documentclass[runningheads]{llncs}
\usepackage{times}
\usepackage{xcolor}
\usepackage{soul}
\usepackage[utf8]{inputenc}
\usepackage[small]{caption}

\usepackage{graphicx}  
\usepackage{wrapfig}

\usepackage{xspace} 
\usepackage{subfigure}
\usepackage{float}
\usepackage{amsmath}
\usepackage{amsfonts}
\usepackage[flushleft]{threeparttable}

\usepackage{algorithm,algorithmicx,algpseudocode}
\usepackage{booktabs}
\usepackage{multirow}

\newcommand{\transRec}{\textsc{TransRev}\xspace}
\newcommand{\svd}{\textsc{SVD}\xspace}
\newcommand{\nmf}{\textsc{NMF}\xspace}
\newcommand{\hft}{\textsc{HFT}\xspace}
\newcommand{\jmars}{\textsc{JMARS}\xspace}

\newcommand{\hide}[1]{}

\begin{document}
\title{\textsc{TransRev}: Modeling Reviews as Translations from Users to Items}
%
%

\author{
Alberto Garc\'ia-Dur\'an\inst{1}
  \and
  Roberto Gonz\'alez\inst{1}
  \and
  Daniel O\~noro-Rubio\inst{1}
  \and
  Mathias Niepert\inst{1}
  \and
  Hui Li\inst{2}\thanks{Work done while interning at NEC Labs Europe}
}
\authorrunning{A. Garc\'ia-Dur\'an et al.}
%
\institute{NEC Labs Europe \\
\email{\string{alberto.duran, roberto.gonzalez, daniel.onoro, mathias.niepert\string}@neclab.eu}\\
\and
The University of Hong Kong \\
\email{hli2@cs.hku.hk}}
\maketitle              
\begin{abstract}
The text of a review expresses the sentiment a customer has towards a particular product. This is exploited in sentiment analysis where  machine learning models are used to predict the review score from the text of the review. Furthermore, the products costumers have purchased in the past are indicative of the products they will purchase in the future. This is what recommender systems exploit by learning models from purchase information to predict the items a customer might be interested in. We propose \textsc{TransRev}, an approach to the product recommendation problem that integrates ideas from recommender systems, sentiment analysis, and multi-relational learning into a joint learning objective. 

\textsc{TransRev} learns vector representations for users, items, and reviews. The embedding of a review is learned such that (a) it performs well as input feature of a regression model for sentiment prediction; and (b) it always translates the reviewer embedding to the embedding of the reviewed items. This allows \textsc{TransRev} to approximate a review embedding at test time as the difference of the embedding of each item and the user embedding. The approximated review embedding is then used with the regression model to predict the review score for each item. \textsc{TransRev} outperforms state of the art recommender systems on a large number of benchmark data sets. Moreover, it is able to retrieve, for each user and item, the review text from the training set whose embedding is most similar to the approximated review embedding.
\end{abstract}
\section{Introduction}
\label{intro}
Online retail is a growing market with sales accounting for \$394.9 billion or 11.7\% of total US retail sales in 2016. In the same year, e-commerce sales accounted for 41.6 percent of all retail sales growth.
For some entertainment products such as movies, books, and music, online retailers have long outperformed traditional in-store retailers. 
One of the driving forces of this success is the ability of online retailers to collect purchase histories of customers, online shopping behavior, and reviews of products for a very large number of users. This data is driving several machine learning applications in online retail, of which personalized recommendation is the most important one. With recommender systems online retailers can provide personalized product recommendations and anticipate purchasing behavior.


\begin{figure}[t!]
\fbox{
\begin{minipage}{.5\textwidth}
\centering
\includegraphics[width=0.8\textwidth]{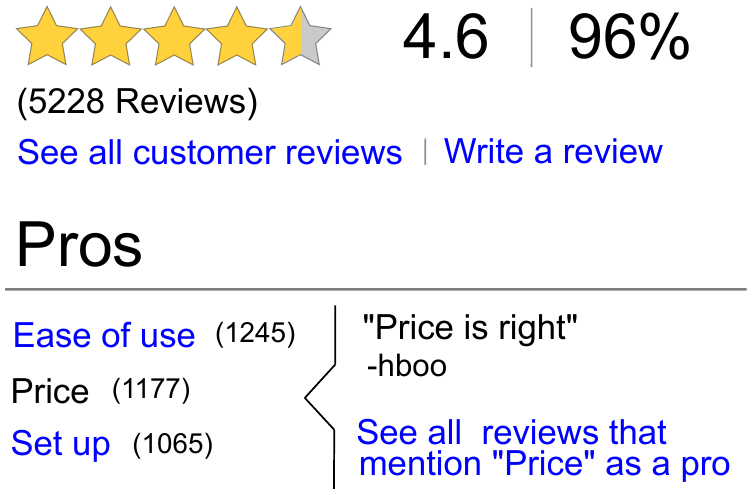}
\end{minipage}\hspace{-5mm}
\begin{minipage}{.5\textwidth}
\centering
\includegraphics[width=0.7\textwidth]{user_graph.pdf}
\end{minipage}
}
\caption{\label{fig:example} (Left) A typical product summary with review score and `Pros'. (Right) A small bipartite graph modeling customers (users), products (items), product reviews, and review scores. }
\end{figure}

In addition, the availability of product reviews allows users to make more informed purchasing choices and companies to analyze costumer sentiment towards their products. The latter was coined sentiment analysis and is concerned with machine learning approaches that map written text to scores. Nevertheless, even the best sentiment analysis methods cannot help in determining which \emph{new} products a costumer might be interested in.  The obvious reason is that costumer reviews are not available for products they have not purchased yet.

In recent years the availability of large corpora of product reviews has driven text-based research in the recommender system community (e.g. \cite{mcauley2013hidden,ling2014ratings,bao2014topicmf}). Some of these novel methods extend latent factor models to leverage review text by employing an explicit mapping from text to either user or item factors. At prediction time, these models predict product ratings based on some operation (typically the dot product) applied to the user and product representations. Sentiment analysis, however, is usually applied to some representation (e.g. bag-of-words) of review text but in a recommender system scenario the review is not available at prediction time. 

With this paper we propose \textsc{TransRev}, a method that combines a personalized recommendation learning objective with a sentiment analysis objective into a joint learning objective. \textsc{TransRev} learns vector representations for users, items, and reviews jointly. The crucial advantage of \textsc{TransRev} is that the review embedding is learned such that it corresponds to a translation that moves the embedding of the reviewing user to the embedding of the item the review is about. This allows \textsc{TransRev} to approximate a review embedding at test time as the difference of the item and user embedding despite the absence of a review from the user for that item. The approximated review embedding is then used in the sentiment analysis model to predict the review score. 
Moreover, the approximated review embedding can be used to retrieve reviews in the training set deemed most similar by a distance measure in the embedding space. These retrieved reviews could be used for several purposes. For instance, such reviews could be provided to users as a starting point for a review, lowering the barrier to writing reviews.  

\hide{
 \begin{figure}[t!]
 \centering
 \fbox{\includegraphics[width=0.32\textwidth]{user_graph.pdf}}
 \caption{\label{fig:data} A small bipartite graph modeling customers (users), products (items), product reviews, and review scores. }
 \end{figure}
}

We performed an extensive set of experiments to evaluate the performance of \textsc{TransRev} on standard recommender system data sets. \textsc{TransRev} outperforms state of the art methods on 15 of the 19 data sets. 
Moreover, we qualitatively compare actual reviews with the retrieved ones by \textsc{TransRev} based on a similarity metric in the review embedding space. Finally, we discuss some weaknesses of \textsc{TransRev} and possible future research directions.



\section{\transRec: Modeling Reviews as Translations in Vector Space}
\label{transRec}

We address the problem of learning prediction models for the product recommendation problem.
There are a set of users $\mathbf{U}$, a set of items $\mathbf{I}$, and a set of reviews $\mathbf{R}$. Each $\mathtt{rev}_{(\mathtt{u,i})} \in \mathbf{R}$ represents a review written by user $\mathtt{u}$ for item
$\mathtt{i}$. Hence, $\mathtt{rev}_{(\mathtt{u,i})} = [\mathtt{t}_1, \cdots, \mathtt{t}_{n}]$, that is, each review is a sequence of $n$ tokens. In the following we refer to $(\mathtt{u}, \mathtt{rev}_{(\mathtt{u,i})}, \mathtt{i})$ as a \emph{triple}. Each such triple is associated with the review score $\mathtt{r}_{(\mathtt{u,i})}$ given by the user $\mathtt{u}$ to item $\mathtt{i}$. 



\transRec embeds all users, items and reviews into a latent space where the embedding of a user plus the embedding of the review is learned to be close to the embedding of the reviewed item. It simultaneously learns a regression model to predict the rating given a review text. At prediction time, reviews are not available, but the modeling assumption of \transRec allows to predict the review embedding by taking the difference of the embedding of the item and user. Then this approximation is used as input feature of the regression model to perform rating prediction.

\begin{figure*}[t!]
\centering
\includegraphics[width=1.0\textwidth]{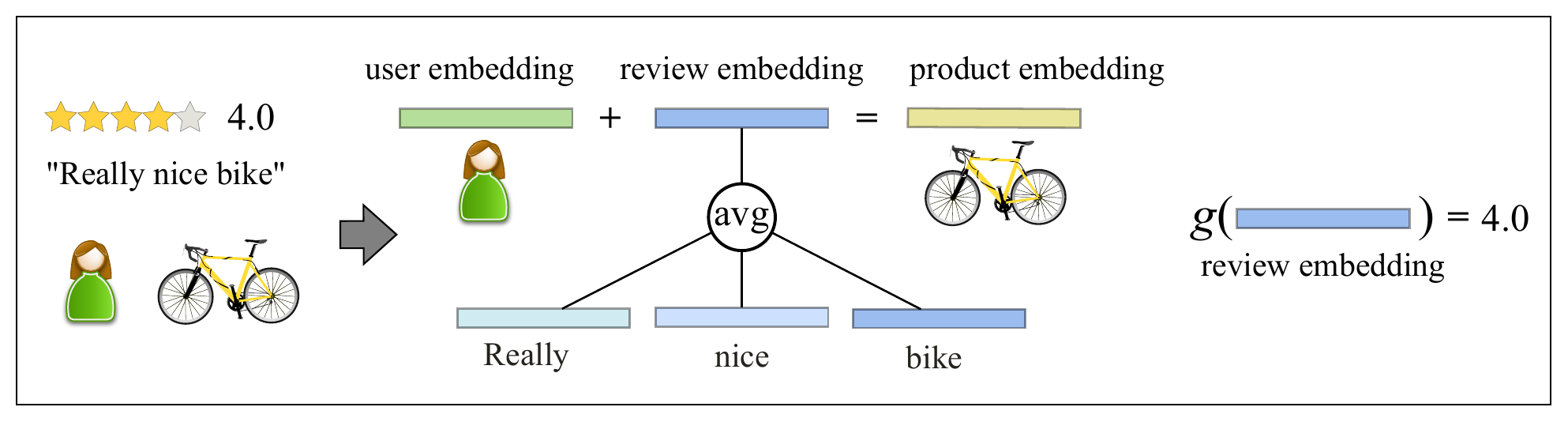}
\caption{\label{fig:TransRec} At training time, a function's parameters are learned to compute the review embedding from the word token embeddings such  that the embedding of the user translated by the review embedding is similar to the product embedding. At the same time, a regression model $g$ is trained to perform well on predicting ratings.}
\end{figure*}

\begin{figure*}[t!]
\centering
\includegraphics[width=0.58\textwidth]{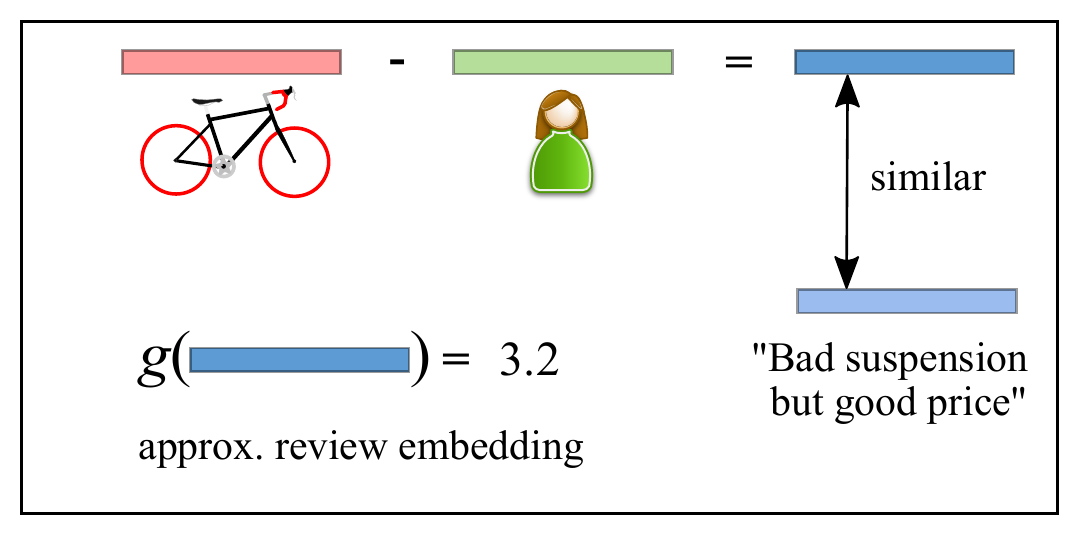}
\caption{\label{fig:TransRec-testing} At test time, the review embedding is approximated as the difference between the product and user embeddings. The approximated review embedding is used to predict the rating and to retrieve similar reviews. }
\end{figure*}

\transRec embeds all nodes and reviews into a latent space $\mathbb{R}^k$ ($k$ is a model hyperparameter). The review embeddings are computed by applying a learnable function $f$ to the token sequence of the review
$$\mathbf{h}_{\mathtt{rev_{(u,i)}}} = f(\mathtt{rev_{(u,i)}}).$$
The function $f$ can be parameterized (typically with a neural network such as a recursive or convolutional neural network) but it can also be a simple parameter-free aggregation function that computes, for instance, the element-wise average or maximum of the token embeddings. 

We propose and evaluate a simple instance of $f$ where the review embedding $\mathbf{h}_{\mathtt{rev}_{\mathtt{(u,i)}}}$ is the average of the embeddings of the tokens occurring in the review. More formally, 
\begin{equation}
\mathbf{h}_{\mathtt{rev}_{\mathtt{(u,i)}}} = f(\mathtt{rev}_{\mathtt{(u,i)}}) = \cfrac{1}{|\mathtt{rev}_{\mathtt{(u,i)}}|} \sum_{\mathtt{t} \in \mathtt{rev}_{\mathtt{(u,i)}}} \mathbf{v}_{\mathtt{t}} + \mathbf{h}_0,
\end{equation}
where $\mathbf{v}_{\mathtt{t}}$ is the embedding associated with token $\mathtt{t}$ and $\mathbf{h}_0$ is a review bias which is common to all reviews and takes values in $\mathbb{R}^k$. The review bias is of importance since there are some reviews all of whose tokens are not in the training vocabulary. In these cases we have $\mathbf{h}_{\mathtt{rev}_{(\mathtt{u,i})}} = \mathbf{h}_0$.

The learning of the item, review, and user embeddings is determined by two learning objectives. The first objective guides the joint learning of the parameters of the regression model and the review embeddings such that the regression model performs well at review score prediction


\begin{equation}
\label{regressor}
\min \mathcal{L}_{1} = \min  \sum_{((\mathtt{u},\mathtt{rev}_{(\mathtt{u,i})},\mathtt{i}),\mathtt{r}_{\mathtt{(u,i)}})\in S} \left(g(\mathbf{h}_{\mathtt{rev_{\mathtt{(u,i)}}}}\right) - \mathtt{r}_{\mathtt{(u,i)}})^{2},
\end{equation}
where $S$ is the set of training triples and their associated ratings, and $g$ is a learnable regression function $\mathbb{R}^k \rightarrow \mathbb{R}$ that is applied to the representation of the review $\mathbf{h}_{\mathtt{rev_{\mathtt{(u,i)}}}}$.

While $g$ can be an arbitrary complex function, the instance of $g$ used in this work is as follows

\begin{equation}
\label{regressorInstance}
g(\mathbf{h}_{\mathtt{rev_{\mathtt{(u,i)}}}}) =  \sigma(\mathbf{h}_{\mathtt{rev_{\mathtt{(u,i)}}}}) \mathbf{w}^{T} + \mathtt{b}_{\mathtt{(u,i)}},
\end{equation}
where $\mathbf{w}$ are the learnable weights of the linear regressor, $\sigma$ is the sigmoid function $\sigma (x) = \cfrac{1}{1 + e^{-x}}$, and $\mathtt{b}_{\mathtt{(u,i)}}$ is the shortcut we use to refer to the sum of the bias terms, namely the user, item and overall bias: $\mathtt{b}_{(\mathtt{u,i})} = \mathtt{b}_{\mathtt{u}} + \mathtt{b}_{\mathtt{i}} + \mathtt{b}_0$.

Of course, in a real-world scenario a recommender system makes rating predictions on items that users have \emph{not rated yet} and, consequently, reviews are not available for those items. The application of the linear regressor of Equation (\ref{regressor}) to new examples, therefore, is not possible at test time. Our second learning procedure aims at overcoming this limitation by leveraging ideas from embedding-based knowledge base completion methods. We want to be able to approximate a review embedding at test time such that this review embedding can be used in conjunction with the learned regression model.  Hence, in addition to the learning objective (\ref{regressor}), we introduce a second objective  that forces  the embedding of a review to be close to the difference between the item and user embeddings. This translation-based modeling assumption is followed in \textsc{TransE} \cite{bordes2013translating} and several other knowledge base completion methods~\cite{guu2015traversing,garcia2015composing}. We include a second term in the objective that drives the distance between (a) the user embedding translated by the review embedding and  (b) the embedding of the item to be small
\begin{equation}
\label{modeling}
\min \mathcal{L}_2 = \min \sum_{((\mathtt{u},\mathtt{rev}_{(\mathtt{u,i})},\mathtt{i}),\mathtt{r}_{\mathtt{(u,i)}})\in S} || \mathbf{e}_{\mathtt{u}} + \mathbf{h}_{\mathtt{rev}_{\mathtt{(u,i)}}} - \mathbf{e}_{\mathtt{i}} ||_2,
\end{equation}
where $\mathbf{e}_{\mathtt{u}}$ and $\mathbf{e}_{\mathtt{i}}$ are the embeddings of the user and item, respectively.
In the knowledge base embedding literature (cf. \cite{bordes2013translating}) it is common the representations are learned via a margin-based loss, where the embeddings are updated if the score (the negative distance) of a positive triple (e.g. $\mathtt{(Berlin, located}\_\mathtt{in, Germany)})$ is not larger than the score of a negative triple (e.g. $\mathtt{(Berlin, located}\_\mathtt{in, Portugal)})$ plus a margin. Note that this type of learning is required to avoid trivial solutions. The minimization problem of Equation~(\ref{modeling}) can easily be solved by setting $\mathbf{e}_{\mathtt{u}} = \mathbf{h}_{\mathtt{rev}_{\mathtt{(u,i)}}} = \mathbf{e}_{\mathtt{i}} = \mathbf{0}$ $\forall \mathtt{u, i}$. However, this kind of trivial solutions is avoided by jointly optimizing Equations (\ref{regressor}) and (\ref{modeling}), since a degenerate solution like the aforementioned one would lead to a high error with respect to the regression objective (Equation (\ref{regressor})). The overall objective can now be written as
\begin{equation}
\label{model:transrev}
\min_{\Theta} \mathcal{L} = \min_{\Theta} (\mathcal{L}_1 + \lambda \mathcal{L}_2 + \mu ||\Theta||_2),
\end{equation}
where $\lambda$ is a term that weights the approximation loss due to the modeling assumption formalized in Equation~(\ref{modeling}). In our model, $\Theta$ corresponds to the parameters $\mathbf{w}$, $\mathbf{e}$, $\mathbf{v}$, $\mathbf{h}_0$ $\in \mathbb{R}^k$ and the bias terms $\mathtt{b}$.

At test time, we can now approximate review embeddings of $\mathtt{(u,i)}$ pairs \emph{not seen} during training by computing 
$$\boxed{\hat{\mathbf{h}}_{\mathtt{rev}_{\mathtt{(u,i)}}} = \mathbf{e}_i - \mathbf{e}_{\mathtt{u}}.}$$ 
With the trained regression model $g$ we can make rating predictions $\hat{\mathtt{r}}_{\mathtt{(u,i)}}$ for \emph{unseen} $\mathtt{(u, i)}$ pairs by computing
\begin{equation}
\hat{\mathtt{r}}_{\mathtt{(u,i)}} = g(\hat{\mathbf{h}}_{\mathtt{rev}_{\mathtt{u,i}}}).
\end{equation}
Contrary to training, now the regression model $g$ is applied over $\hat{\mathbf{h}}_{\mathtt{rev}_{\mathtt{u,i}}}$, instead of $\mathbf{h}_{\mathtt{rev}_{\mathtt{u,i}}}$, which is not available at test time.

All parameters of the parts of the objective are jointly learned with stochastic gradient descent. More details regarding the parameter learning are contained in Section \ref{setting}. Algorithm \ref{alg:approach} illustrates the generic working of \transRec.

\begin{algorithm}[t!]
\caption{{\small Learning \transRec}}
\label{alg:approach}
\begin{algorithmic}[1]
   \State {\bfseries Input:} Training set $S_{0} = \{((\mathtt{u},\mathtt{rev}_{(\mathtt{u,i})},\mathtt{i}), \mathtt{r}_{(\mathtt{u,i})})\}$, embedding dimension $k$, batch size $\mathtt{batchSize}$, number of epochs $\mathtt{numEpochs}$, weight $\lambda$ , regularization term $\mu$. \\
   \State {\bfseries Initialize:} All parameters $\Theta$ are initialized following \cite{glorot2010understanding}.
   \State {$\mathtt{numBatches} = |S_{0}| / \mathtt{batchSize}$}.
   \For{$n=1$ {\bfseries to} $\mathtt{numEpochs}$}
   \State {$S = S_0$}
   \For{$m=1$ {\bfseries to} $\mathtt{numBatches}$}
	\State {$S_{\mathtt{batch}} \gets $Sample$(S,\mathtt{batchSize})$} \Comment{Sampling without replacement}
   \State Update parameters $\Theta$ w.r.t. Equation (\ref{model:transrev}) computed on $S_{\mathtt{batch}}$.
   \EndFor
   \EndFor
\end{algorithmic}
\end{algorithm}

\section{Related Work}
\label{related}

There are three lines of research related to our work. Recommender systems, sentiment analysis and multi-relational graph completion. There is an extensive body of work on recommender systems \cite{allen1990user,breese1998empirical,rennie2005fast,sarwar2001item,brun2010towards,wang2015collaborative,guo2015trustsvd,dong2017hybrid}. Singular Value Decomposition (\svd) \cite{koren2009matrix} computes the review score prediction as the dot product between the item embeddings and the user embeddings plus some learnable bias terms. Due to its simplicity and performance on numerous data sets it is still one of the most used methods for product recommendations. Even though there has been a flurry of research on predicting ratings from the interaction of latent representations of users and items, there is not much work on incorporating review text despite its availability in several corpora. \cite{jakob2009beyond} was one of the first approaches that demonstrated that features extracted from review text are are useful in learned models to improve the accuracy of rating predictions. 
Most of the previous research that explored the utility of review text for rating prediction can be classified into two categories.
\begin{itemize}
\item \textbf{Semi-supervised approaches.} \hft \cite{mcauley2013hidden} was one of the first methods combining a supervised learning objective to predict ratings with an unsupervised learning objective (e.g. latent Dirichlet allocation) for text content to regularize the parameters of the supervised model. The idea of combining two learning objectives has been explored in several additional approaches~\cite{ling2014ratings,bao2014topicmf,diao2014jointly,almahairi2015learning}. The methods differ in the unsupervised objectives, some of which are tailored to a specific domain. For example, \jmars \cite{diao2014jointly} outperforms \hft on a movie recommendation data set but it is outperformed by \hft on data sets similar to those used in our work~\cite{wu2016explaining}.
\item \textbf{Supervised approaches.} Methods that fall into this category such as \cite{seo2017representation,zheng2017joint,catherine2017transnets} learn latent representations of users and items from the text content so as to perform well at rating prediction. The learning of the latent representations is done via a deep architecture. The approaches differences lie mainly in the neural architectures they employ. 
\end{itemize}


There is one crucial difference between the aforementioned methods and \transRec. \transRec predicts the review score based on an approximation of the review embedding computed at test time. Moreover, since \transRec is able to approximate a review embedding, we can use this embedding to retrieve reviews in the training set deemed most similar by a distance metric in the embedding space.   

Similar to sentiment analysis methods, \textsc{TransRev} trains a regression model that predicts the review rating from the review text. Contrary to the typical setting in which sentiment analysis methods operate, however, review text is not available at prediction time in the recommender system setting. Consequently, the application of sentiment analysis for recommender systems is not directly possible. In the simplest case, a sentiment analysis method is a linear regressor applied to a text embedding (Equation (\ref{regressorInstance})). \transRec trains such a regression model to perform well in conjunction with the approximated review embedding. 

The third research theme related to \transRec is knowledge base completion. In the last years, many embedding-based methods have been proposed to infer missing relations in knowledge bases based on function that computes a likelihood score based on the embeddings of entities and relation types. Due to its simplicity and good performance, there is a large body of work on translation-based scoring functions~\cite{bordes2013translating,guu2015traversing,garcia2015composing}. \cite{mcauley2017} propose an approach to large-scale sequential sales prediction that embeds items into a transition space where user embeddings are modeled as translation vectors operating on item sequences. The associated optimization problem is formulated as a sequential Bayesian ranking problem~\cite{rendle2010factorizing}. To the best of our knowledge, \cite{mcauley2017} is the first work in leveraging ideas from knowledge base completion methods for recommender system. Whereas \transRec addresses the problem of rating prediction by incorporating review text, \cite{mcauley2017} addresses the different problem of sequential recommendation. Therefore the experimental comparison to that work is not possible. In \transRec the review embedding translates the user embedding to the product embedding. In \cite{mcauley2017}, the user embedding translates a product embedding to the embedding of the next purchased product.  
\transRec is also novel in that the approximated review embeddings can be used to retrieve, from an existing training set, the reviews deemed most similar by a distance metric in the embedding space.

\section{Experimental Setup}
\label{exps}

We conduct several experiments to empirically compare \transRec to state of the art methods for product recommendation. More specifically, we compare \transRec to competitive matrix factorization methods as well as methods that take advantage of review text.  Moreover, we provide some qualitative results on retrieving training reviews most similar to the approximated reviews at test time.

\subsection{Data Sets}
\label{data}
We evaluate the various methods on two commonly used data sets. The Yelp Business Rating Prediction Challenge\footnote{https://www.kaggle.com/c/yelp-recsys-2013} data set consists of reviews on restaurants in Phoenix (United States). The  Amazon Product Data\footnote{http://jmcauley.ucsd.edu/data/amazon} has been extensively used in previous works \cite{mcauley2013hidden,mcauley2015inferring,mcauley2015image}. The data set consists of reviews and product metadata from Amazon from May 1996 to July 2014. We focus on the 5-core versions (which contain at least 5 reviews for each user and item) of those data sets. There are 24 product categories from which we have selected those 12 used in \cite{seo2017representation}, plus 6 randomly picked categories out of the 12 remaining ones. We treat each of these resulting 18 data sets independently in our experiments. Ratings in both benchmark data sets are integer values between 1 and 5. As in previous work, we randomly sample 80\% of the reviews as training, 10\% as validation, and 10\% as test data. We remove reviews from the validation and test splits if they involve either a product or a user that is not part of the training data. 

\subsection{Review Text Preprocessing}
\label{preprocess}

We follow the same preprocessing steps for each data set. First, we lowercase the review texts and apply the regular expression ``$\setminus w+$'' to tokenize the text data, discarding those words that appear in less than 0.1$\%$ of the reviews of the data set under consideration. For the Amazon data sets, both full reviews and short summaries (rarely having more than 30 words) are available. Since classifying short documents into their sentiment is less challenging than doing the same for longer text~\cite{bermingham2010classifying}, we have used the reviews summaries for our work. For the Yelp data only full reviews are available. We truncate these reviews to the first 200 words. Some statistics of the preprocessed data sets are summarized in Table \ref{tab:stats-data}.

\begin{table}[t!]
\centering
\scalebox{0.75}{
 \begin{threeparttable} 
\resizebox{\columnwidth}{!}{
\begin{tabular}{c|c|c|c|c|c|c|}
\cline{2-7} 
 & $\#$Users & $\#$Items & $\#$Words & $\#$Training & $\#$Valid. & $\#$Test \\ 
\hline
\multicolumn{1}{|c|}{\multirow{1}{*}{Amazon}} &  \multirow{2}{*}{5,131} & \multirow{2}{*}{1,686}  & \multirow{2}{*}{513} & \multirow{2}{*}{27,610}  & \multirow{2}{*}{3,449}  & \multirow{2}{*}{3,445}\\ 
\multicolumn{1}{|c|}{Instant Video} & & & & & & \\ \hline
\multicolumn{1}{|c|}{Automotive} & 2,929 & 1,836 & 589 & 15,741 & 1,965 & 1,963 \\ \hline
\multicolumn{1}{|c|}{Baby} & 22,364 & 12,102 & 497 & 124,978 & 15,612 & 15,613 \\ \hline
\multicolumn{1}{|c|}{Cds and Vinyl} & 75,259 & 64,444 & 576 & 813,897 & 101,600 & 101,581 \\ \hline
\multicolumn{1}{|c|}{\multirow{1}{*}{Grocery}} & \multirow{2}{*}{14,682} & \multirow{2}{*}{8,714} & \multirow{2}{*}{565} & \multirow{2}{*}{116,192} & \multirow{2}{*}{14,502} & \multirow{2}{*}{14,499} \\
\multicolumn{1}{|c|}{Gourmet Food} & & & & & & \\ \hline
\multicolumn{1}{|c|}{Health} & \multirow{2}{*}{38,610} & \multirow{2}{*}{18,535} & \multirow{2}{*}{573} & \multirow{2}{*}{261,102} & \multirow{2}{*}{32,588} & \multirow{2}{*}{32,585} \\
\multicolumn{1}{|c|}{Personal Care} & & & & & & \\ \hline
\multicolumn{1}{|c|}{Kindle Store}  & 68,224 & 61,935 & 456 & 717,845 & 89,628 & 89,637 \\ \hline
\multicolumn{1}{|c|}{Musical} & \multirow{2}{*}{1,430} & \multirow{2}{*}{901} & \multirow{2}{*}{512} & \multirow{2}{*}{7,925} & \multirow{2}{*}{989} & \multirow{2}{*}{985} \\
\multicolumn{1}{|c|}{Instruments} & & & & & & \\ \hline
\multicolumn{1}{|c|}{Office Products} & 4,906 & 2,421 & 652 & 41,687 & 5,210 & 5,206 \\ \hline
\multicolumn{1}{|c|}{Patio, Lawn} & \multirow{2}{*}{1,687} & \multirow{2}{*}{963} & \multirow{2}{*}{697} & \multirow{2}{*}{10,320} & \multirow{2}{*}{1,279} & \multirow{2}{*}{1,285} \\
\multicolumn{1}{|c|}{Garden} & & & & & & \\ \hline
\multicolumn{1}{|c|}{Pet Supplies} & 19,857 & 8,511 & 515 & 120,831 & 15,073 & 15,070 \\ \hline
\multicolumn{1}{|c|}{Tools} & \multirow{2}{*}{16,639} & \multirow{2}{*}{10,218} & \multirow{2}{*}{587} & \multirow{2}{*}{103,373} & \multirow{2}{*}{12,911} & \multirow{2}{*}{12,910} \\
\multicolumn{1}{|c|}{Home Improv.} & & & & & & \\ \hline
\multicolumn{1}{|c|}{Toys}  & \multirow{2}{*}{19,413} & \multirow{2}{*}{11,925} & \multirow{2}{*}{516} & \multirow{2}{*}{127,712}  & \multirow{2}{*}{15,864} & \multirow{2}{*}{15,850} \\ 
\multicolumn{1}{|c|}{Games} & & & & & & \\ \hline
\multicolumn{1}{|c|}{Beauty} & 22,364 & 12,102 & 497 & 150,452 & 18,774 & 18,783 \\ \hline
\multicolumn{1}{|c|}{Digital Music} & 5,542 & 3,569  & 625 & 48,283 & 6,029 & 6,021\\ \hline
\multicolumn{1}{|c|}{Video Games} & 24,304 & 10,673 & 591 & 175,650 & 21,948 & 21,937\\ \hline
\multicolumn{1}{|c|}{Sports} & \multirow{2}{*}{35,599} & \multirow{2}{*}{18,358} & \multirow{2}{*}{530} & \multirow{2}{*}{224,596} & \multirow{2}{*}{28,045} & \multirow{2}{*}{28,035} \\
\multicolumn{1}{|c|}{Outdoors} & & & & & & \\ \hline
\multicolumn{1}{|c|}{Cell Phones} & \multirow{2}{*}{27,880} & \multirow{2}{*}{10,430} & \multirow{2}{*}{504} & \multirow{2}{*}{149,668} & \multirow{2}{*}{18,667} & \multirow{2}{*}{18,673} \\ 
\multicolumn{1}{|c|}{Accesories} & & & & & & \\ \hline
\hline
\multicolumn{1}{|c|}{\textsc{Yelp}} & 45,981 & 11,538 & 5,314 & 183,886 &  20,315 &  20,294 \\
\hline
\end{tabular}
 }
  \end{threeparttable}
  }
  \caption{\label{tab:stats-data} Statistics of the data sets used for the experimental evaluation. }
\end{table}

\begin{table*}[t!]
\centering
\scalebox{1}{ 
  \begin{threeparttable} 
\begin{tabular}{l|c|c|c|c|c|c|c|}
\cline{2-8} 
 & Offset & Attn+CNN & \nmf & \svd & \hft & DeepCoNN & \transRec  \\ 
\hline
\multicolumn{1}{|l|}{Amazon Instant Video}  &  1.180 & 0.936  & 0.946 & 0.904 & 0.888 & 0.943 & \textbf{0.884}  \\ 
\multicolumn{1}{|l|}{Automotive} & 0.948 & 0.881 & 0.876 & 0.857 & 0.862 & \textbf{0.753} & 0.855   \\
\multicolumn{1}{|l|}{Baby} & 1.262 & 1.176 & 1.171 & 1.108 & 1.104 & 1.154 & \textbf{1.100}  \\
\multicolumn{1}{|l|}{Cds and Vinyl} & 1.127 & 0.866 & 0.871 & 0.863 & \textbf{0.854} & 0.888 & \textbf{0.854}  \\
\multicolumn{1}{|l|}{Grocery and Gourmet Food} & 1.165 & 1.004 & 0.985 & 0.964 & 0.961 & 0.973 & \textbf{0.957} \\
\multicolumn{1}{|l|}{Health and Personal Care} & 1.200 & 1.054 & 1.048 & 1.016 & 1.014 & 1.081 & \textbf{1.011} \\
\multicolumn{1}{|l|}{Kindle Store}  & 0.87 & 0.617 & 0.624 & 0.607 & \textbf{0.593} & 0.648 & 0.599 \\ 
\multicolumn{1}{|l|}{Musical Instruments} & 0.733 & 0.703 & 0.725 & 0.694 & 0.692 & 0.723 & \textbf{0.690}  \\
\multicolumn{1}{|l|}{Office Products} & 0.876 & 0.726 & 0.742 & 0.727 & 0.727 & 0.738 & \textbf{0.724} \\
\multicolumn{1}{|l|}{Patio, Lawn and Garden} & 1.156 & 0.999 & 0.958 & 0.950 & 0.956 & 1.070 & \textbf{0.941}  \\
\multicolumn{1}{|l|}{Pet Supplies} & 1.354 & 1.236 & 1.241 & 1.198 & 1.194 & 1.281 & \textbf{1.191} \\
\multicolumn{1}{|l|}{Tools and Home Improvement} & 1.017 & 0.938 & 0.908 & 0.884 & 0.884 & 0.946 & \textbf{0.879} \\
\multicolumn{1}{|l|}{Toys and Games}  & 0.975 & - & 0.821 & 0.788  & \textbf{0.784} & 0.851 & \textbf{0.784} \\ 
\multicolumn{1}{|l|}{Beauty} & 1.322 & - & 1.204 & 1.168 & 1.165 & 1.184 & \textbf{1.158} \\
\multicolumn{1}{|l|}{Digital Music} & 1.137 & -  & 0.805 & 0.797 & 0.793 & 0.835 & \textbf{0.782}  \\
\multicolumn{1}{|l|}{Video Games} & 1.401 & - & 1.138 & 1.093 & 1.086 & 1.133 & \textbf{1.082} \\
\multicolumn{1}{|l|}{Sports and Outdoors} & 0.931 & - & 0.856 & 0.828 & 0.824 & 0.882 & \textbf{0.823} \\
\multicolumn{1}{|l|}{Cell Phones and Accesories} & 1.455 & - & 1.357 & 1.290 & 1.285 & 1.365 & \textbf{1.279} \\
\cline{1-1} & $1.117^*$ & - & $0.959^*$ & $0.930^*$ & $0.926^*$ & $0.969^*$ & \textbf{0.921$^*$} \\
\cline{2-8} \addlinespace[1mm]
\hline 
\multicolumn{1}{|l|}{\textsc{Yelp}} & 1.385 & 1.212 & 1.229 & 1.158 &  1.148 & 1.215 & \textbf{1.144} \\
\hline
\end{tabular}
  \end{threeparttable}
  }
  \caption{\label{tab:exp-table}  MSE for  \transRec and  state-of-the-art approaches. $^*$ indicates the macro MSE across the Amazon data sets.}
\end{table*}
\normalsize

\subsection{Baselines}
\label{baselines}

We compare to the matrix factorization-based methods \svd and \nmf (non-negative matrix factorization) as well as approaches that leverage review text for rating prediction in a semi-supervised manner like \hft, and in a supervised manner such as \textsc{Attn+CNN}~\cite{seo2017representation,SeoHYL17} and \textsc{DeepCoNN}~\cite{zheng2017joint}. We also compare to a simple baseline \textsc{Offset} that simply uses the average rating in the training set as the prediction.

\begin{table}[t!]
\centering
\resizebox{0.5\columnwidth}{!}{
\begin{tabular}{ccccc}
\hline
\textit{k} & Baby  & \begin{tabular}[c]{@{}c@{}}Digital \\ Music\end{tabular} & \begin{tabular}[c]{@{}c@{}}Office \\ Products\end{tabular} & \begin{tabular}[c]{@{}c@{}}Tools\&Home \\ Improv.\end{tabular} \\ \hline
4          & 1.100 & 0.782                                                    & 0.724                                                      & 0.880                                                          \\
8          & 1.100 & 0.782                                                    & 0.723                                                      & 0.878                                                          \\
16         & 1.100 & 0.782                                                    & 0.724                                                      & 0.879                                                          \\
32         & 1.102 & 0.785                                                    & 0.722                                                      & 0.888                                                          \\
64         & 1.099 & 0.787                                                    & 0.726                                                      & 0.888                                                          \\ \hline
\end{tabular}
}
\caption{\label{tab:dimensionality} Effect of the latent dimension $k$ in \transRec.}
\end{table}

\subsection{Parameter Setting}
\label{setting}

We set the dimension $k$ of the embedding space to $16$ for all methods. We evaluated the robustness of \transRec to changes in the hyper-parameter $k$ but did not observe any significant performance difference. This is in line with previous work on the Yelp and Amazon data sets that observed that \hft and \svd did not show any improvements for $k > 10$~\cite{mcauley2013hidden}. For \svd and \nmf we used the Python package \textsc{SurPRISE}\footnote{https://pypi.python.org/pypi/scikit-surprise}, whose optimization is performed by vanilla stochastic gradient descent, and chose the learning rate and regularization term on the validation set from the values $[0.001, 0.005, 0.01, 0.05, 0.1]$ and $[0.00001, 0.00005, 0.0001, 0.0005, 0.001]$. For \hft we used the original implementation of the authors\footnote{http://cseweb.ucsd.edu/~jmcauley/code/code$\_$RecSys13.tar.gz} and validated the regularization term from the values $[0.001, 0.01, 0.1, 1, 10, 50]$. For \transRec we validated $\lambda$ among the values $[0.1, 0.25, 0.5, 1]$ and the learning rate of the optimizer and regularization term ($\mu$ in our model) from the same set of values as for \svd and \nmf. To ensure a fair comparison with \svd and \nmf, we also use vanilla SGD to optimize \transRec. \transRec's parameters were randomly initialized \cite{glorot2010understanding}. Parameters for \hft were learned with L-BFGS which was run for 2,500 learning iterations and validated every 50 iterations.

\begin{figure*}[t!]
\centering
\includegraphics[width=0.75\textwidth]{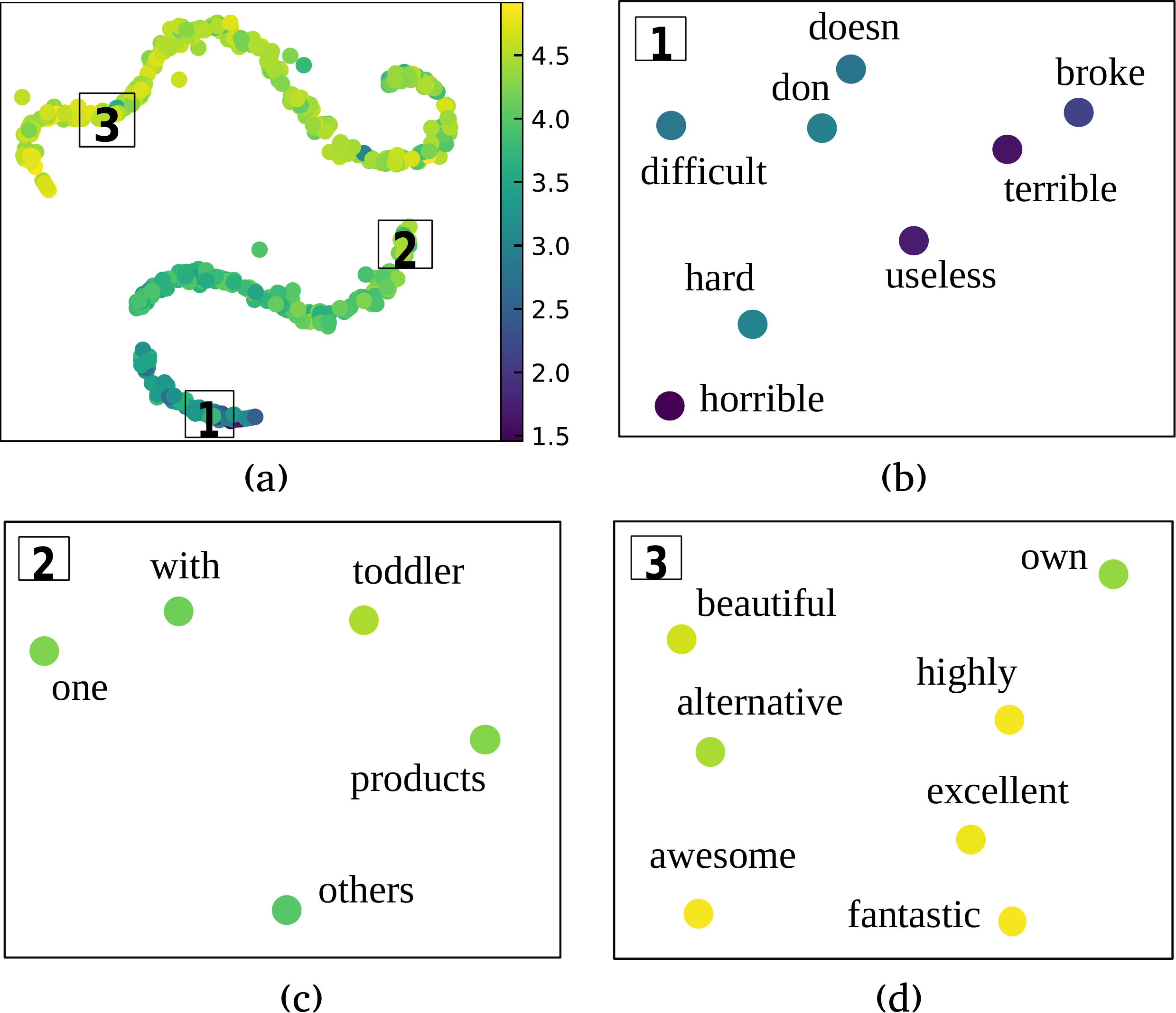}
\caption{\label{fig:visualization} a) Two-dimensional t-SNE representations of the word embeddings learned by \transRec for the \textit{Beauty} data set. The color bar represents the average rating of the reviews where each word appears. b), c) and d) depict regions of the embedding space where negative, neutral and positive words are clustered, respectively.}
\end{figure*}

A single learning iteration performs SGD with all review triples in the training data and their associated ratings. For \transRec we used a batch size of 64. We ran \svd, \nmf and \transRec for a maximum of 500 epochs and validated every $10$ epochs. All methods are validated according to the Mean Squared Error (MSE)
$$\mathtt{MSE} = \cfrac{1}{|\mathcal{T}|} \sum_{(\mathtt{u,i})\in \mathcal{T}} ( \mathtt{r}_{\mathtt{u,i}} - \hat{\mathtt{r}}_{\mathtt{u,i}} )^2,$$
where $\mathcal{T}$ is either the validation or test set.
The implementation of Attn+CNN is not publicly available, so we directly copied the MSE from \cite{seo2017representation} where the  training, validation, and test data sets have the same proportions ($80\%, 10\%, 10\%$). For \textsc{DeepCoNN} the original author code is not available and we used a third-party implementation\footnote{https://github.com/chenchongthu/DeepCoNN}. We applied the default hyperparameters values for dropout and L2 regularization and used the same embedding dimension as for all other methods.

\hide{
\transRec was implemented with the Theano \cite{bergstra2010theano} wrapper Keras \cite{cholletkeras}. All experiments were run on commodity hardware with 128GB RAM, a single 2.8 GHz CPU, and a TitanX GPU.
}

\subsection{Sensitivity}
\label{sensitivity}

We randomly selected the 4 data sets \textit{Baby, Digital Music, Office} and \textit{Tools}$\&$\textit{Home Improvement} from the Amazon data and evaluated different values of $k$ for user, item and word embedding sizes. We increase $k$ from 4 to 64 and list the MSE scores in Table~\ref{tab:dimensionality}. We only observe insignificant differences in the corresponding model's performances. This observation is in line with \cite{mcauley2013hidden}.

\subsection{Results}
\label{results}

The experimental results are listed in Table~\ref{tab:exp-table} where the best performance is in bold font. \transRec achieves the best performance on 17 out of the 19 data sets. In line with previous work~\cite{jakob2009beyond,mcauley2013hidden}, both \transRec and \hft  outperform methods that do not take advantage of review text. \transRec is competitive with and often outperforms \hft on the benchmark data sets under consideration. To quantify that the rating predictions made by \hft and \transRec are significantly different we have computed the dependent t-test for paired samples and for all data sets where \transRec outperforms \hft, the p-value is smaller than 0.01.

We only copied the numbers of Attn+CNN from \cite{seo2017representation} since an implementation is not available. This could lead to differences in the results due to the different randomly sampled training, validation, and test sets. However, in addition to the results in this paper, Attn+CNN was compared to some of the baselines in related work~ \cite{seo2017representation}. The authors there showed that Attn+CNN performs worse than either \svd or \hft or both in 10 of 12 Amazon data sets. At the same time, in our experiments, \transRec performs better than \hft and \svd on the same data sets with the exception of the Kindle Store category.

\begin{table}[]
\centering
\scalebox{1}{
\begin{tabular}{|l|l|}
\hline
& \\
\multicolumn{1}{|c|}{\textbf{Actual test review}}          & \multicolumn{1}{|c|}{\textbf{Closest training review in embedding space}}      \\ \hline
skin improved (5)             & makes your face feel refreshed (5) \\
love it (5)                   & you'll notice the difference (5)   \\
best soap ever (5)            & I'll never change it (5)           \\
it clumps (2)                 & gives me headaches (1)             \\
smells like bug repellent (3) & pantene give it up (2)             \\
fake fake fake do not buy (1) & seems to be harsh on my skin (2)  \\
saved my skin (5)             & not good quality (2)               \\ \hline
another great release from saliva (5)            & can t say enough good things about this cd (5)           \\
a great collection (5)                 & definitive collection (5)             \\
sound nice (3)                  & not his best nor his worst (4)   \\
a complete massacre of an album (2) & some great songs but overall a dissapointment (3)             \\
the very worst best of ever (1)             & overall a pretty big disappointment (2) \\
what a boring moment (1) & overrated but still allright (3)  \\
great cd (5)             & a brilliant van halen debut album (5)               \\ \hline
\end{tabular}
}
\caption{\label{tab:reviews} Reviews retrieved from the \textit{Beauty} (upper) and \textit{Digital Music} (lower) data sets. In parenthesis the ratings associated to the reviews}
\end{table}

\subsection{Visualization of the Word Embeddings}
\label{visual}

Review embeddings learned by \transRec are learned so as to carry information about user ratings (Equation (\ref{regressor})) and  information about the average word embedding of the words in the review text. As a consequence the learned word embeddings are correlated with ratings. To visualize the correlation between words and ratings we proceed as follows. First, we assign a score to each word that is computed by taking the average rating of the reviews that contain the word. Second, we compute a 2-dimensional representation of the words by applying t-SNE \cite{Maaten:2008} to the 16-dimensional word embeddings learned by \transRec. Figure \ref{fig:visualization} depicts these 2-dimensional word embedding vectors learned for the Amazon \textit{Baby} data set. The corresponding rating scores are indicated by the color of the dots.

The clusters we discovered in Figure \ref{fig:visualization} are interpretable. They are meaningful with respect to the score, observing that the bottom cluster is mostly made up of words with negative connotations (e.g. horrible, useless, terrible$\dots$), the middle one of neutral words (e.g. with, products, others$\dots$) and the upper one of words with positive connotations (e.g. awesome, fantastic, excellent$\dots$). This shows \transRec's ability to learn word embeddings that also capture the sentiment of the review.

\subsection{Suggesting Reviews to Users}
One of the characteristics of \transRec is its ability to approximate the review representation at prediction time. This approximation is used to make a rating prediction, but it can also be used to propose a tentative review on which the user can elaborate on. This is related to a number of approaches \cite{zhang2014explicit,lawlor2015opinionated,qureshi2017lit} on explainable recommendations. We think that this can lower the barrier to write reviews. We compute the Euclidean distance between the approximated review embedding $\hat{\mathbf{h}}_{\mathtt{rev}_{(\mathtt{u,i})}}$ and all review embeddings $\mathbf{h}_{\mathtt{rev}_{(\mathtt{u,i})}}$ from the training set. We then retrieve the review text with the most similar review embedding. We investigate the quality of the tentative reviews that \transRec retrieves for the \textit{Beauty} and \textit{Digital Music} data sets. The example reviews listed in Table \ref{tab:reviews} show that while the overall sentiment is correct in most cases, we can also observe the following shortcomings:

\begin{itemize}
\item The function $f$ chosen in our work is invariant to word ordering and, therefore, cannot learn that bigrams such as ``not good" have a negative meaning.
\item Despite matching the overall sentiment, the actual and retrieved review can refer to different aspects of the product (for example, ``it clumps'' and ``gives me headaches'').
\item Reviews can be specific to a single product. A straightforward improvement could be achieved by retrieving only existing reviews for the specific product under consideration.
\end{itemize}

We believe that more sophisticated sentence and paragraph representations might lead to better results in the review retrieval task. Moreover, a promising line of research has to do with learning representations for reviews that are aspect-specific. It would allow users to obtain retrieved reviews that mention specific aspect of products such as ``ease of use'' and ``price.'' We also think that similar ideas can be followed with data modalities other than review text.

\section{Conclusion}

\transRec is a novel approach for product recommendation combining methods and ideas from the areas of matrix factorization-based recommender systems, sentiment analysis, and knowledge graph completion. \transRec achieves state of the art performance on the data sets under consideration and outperforms existing methods in 15 of these data sets. \transRec is learned so as to be able to approximate, at test time, the embedding of the review as the difference of the embedding of the reviewed item and of the reviewing user. The approximated review embedding can be used with a sentiment analysis method to predict the review score.

%
%
%
 \bibliographystyle{splncs04}
 \bibliography{ref}

\begin{thebibliography}{10}
\providecommand{\url}[1]{\texttt{#1}}
\providecommand{\urlprefix}{URL }
\providecommand{\doi}[1]{https://doi.org/#1}

\bibitem{allen1990user}
Allen, R.B.: User models: Theory, method, and practice. International Journal
  of Man-Machine Studies  \textbf{32}(5),  511--543 (1990)

\bibitem{almahairi2015learning}
Almahairi, A., Kastner, K., Cho, K., Courville, A.C.: Learning distributed
  representations from reviews for collaborative filtering. In: RecSys. pp.
  147--154 (2015)

\bibitem{bao2014topicmf}
Bao, Y., Fang, H., Zhang, J.: Topicmf: Simultaneously exploiting ratings and
  reviews for recommendation. In: {AAAI}. pp.~2--8 (2014)

\bibitem{bermingham2010classifying}
Bermingham, A., Smeaton, A.F.: Classifying sentiment in microblogs: is brevity
  an advantage? In: {CIKM}. pp. 1833--1836 (2010)

\bibitem{bordes2013translating}
Bordes, A., Usunier, N., Garc{\'{\i}}a{-}Dur{\'{a}}n, A., Weston, J.,
  Yakhnenko, O.: Translating embeddings for modeling multi-relational data. In:
  {NIPS}. pp. 2787--2795 (2013)

\bibitem{breese1998empirical}
Breese, J.S., Heckerman, D., Kadie, C.M.: Empirical analysis of predictive
  algorithms for collaborative filtering. In: {UAI}. pp. 43--52 (1998)

\bibitem{brun2010towards}
Brun, A., Hamad, A., Buffet, O., Boyer, A.: Towards preference relations in
  recommender systems. In: Preference Learning (PL 2010) ECML/PKDD 2010
  Workshop (2010)

\bibitem{catherine2017transnets}
Catherine, R., Cohen, W.W.: Transnets: Learning to transform for
  recommendation. In: RecSys. pp. 288--296 (2017)

\bibitem{diao2014jointly}
Diao, Q., Qiu, M., Wu, C., Smola, A.J., Jiang, J., Wang, C.: Jointly modeling
  aspects, ratings and sentiments for movie recommendation {(JMARS)}. In:
  {KDD}. pp. 193--202 (2014)

\bibitem{dong2017hybrid}
Dong, X., Yu, L., Wu, Z., Sun, Y., Yuan, L., Zhang, F.: A hybrid collaborative
  filtering model with deep structure for recommender systems. In: {AAAI}. pp.
  1309--1315 (2017)

\bibitem{garcia2015composing}
Garc{\'{\i}}a{-}Dur{\'{a}}n, A., Bordes, A., Usunier, N.: Composing
  relationships with translations. In: {EMNLP}. pp. 286--290. The Association
  for Computational Linguistics (2015)

\bibitem{glorot2010understanding}
Glorot, X., Bengio, Y.: Understanding the difficulty of training deep
  feedforward neural networks. In: {AISTATS}. {JMLR} Proceedings, vol.~9, pp.
  249--256 (2010)

\bibitem{guo2015trustsvd}
Guo, G., Zhang, J., Yorke{-}Smith, N.: Trustsvd: Collaborative filtering with
  both the explicit and implicit influence of user trust and of item ratings.
  In: {AAAI}. pp. 123--129 (2015)

\bibitem{guu2015traversing}
Guu, K., Miller, J., Liang, P.: Traversing knowledge graphs in vector space.
  In: {EMNLP}. pp. 318--327. The Association for Computational Linguistics
  (2015)

\bibitem{mcauley2017}
He, R., Kang, W., McAuley, J.: Translation-based recommendation. In: RecSys.
  pp. 161--169 (2017)

\bibitem{jakob2009beyond}
Jakob, N., Weber, S.H., M{\"u}ller, M.C., Gurevych, I.: Beyond the stars:
  exploiting free-text user reviews to improve the accuracy of movie
  recommendations. In: 1st international CIKM workshop on Topic-sentiment
  analysis for mass opinion. pp. 57--64 (2009)

\bibitem{koren2009matrix}
Koren, Y., Bell, R.M., Volinsky, C.: Matrix factorization techniques for
  recommender systems. {IEEE} Computer  \textbf{42}(8),  30--37 (2009)

\bibitem{lawlor2015opinionated}
Lawlor, A., Muhammad, K., Rafter, R., Smyth, B.: Opinionated explanations for
  recommendation systems. In: International Conference on Innovative Techniques
  and Applications of Artificial Intelligence. pp. 331--344. Springer (2015)

\bibitem{ling2014ratings}
Ling, G., Lyu, M.R., King, I.: Ratings meet reviews, a combined approach to
  recommend. In: RecSys. pp. 105--112 (2014)

\bibitem{Maaten:2008}
Maaten, L.v.d., Hinton, G.: Visualizing data using t-sne. Journal of Machine
  Learning Research  \textbf{9},  2579--2605 (2008)

\bibitem{mcauley2013hidden}
McAuley, J.J., Leskovec, J.: Hidden factors and hidden topics: understanding
  rating dimensions with review text. In: RecSys. pp. 165--172 (2013)

\bibitem{mcauley2015inferring}
McAuley, J.J., Pandey, R., Leskovec, J.: Inferring networks of substitutable
  and complementary products. In: {KDD}. pp. 785--794 (2015)

\bibitem{mcauley2015image}
McAuley, J.J., Targett, C., Shi, Q., van~den Hengel, A.: Image-based
  recommendations on styles and substitutes. In: {SIGIR}. pp. 43--52 (2015)

\bibitem{qureshi2017lit}
Qureshi, M.A., Greene, D.: Lit@ eve: Explainable recommendation based on
  wikipedia concept vectors. In: Joint European Conference on Machine Learning
  and Knowledge Discovery in Databases. pp. 409--413. Springer (2017)

\bibitem{rendle2010factorizing}
Rendle, S., Freudenthaler, C., Schmidt{-}Thieme, L.: Factorizing personalized
  markov chains for next-basket recommendation. In: {WWW}. pp. 811--820 (2010)

\bibitem{rennie2005fast}
Rennie, J.D.M., Srebro, N.: Fast maximum margin matrix factorization for
  collaborative prediction. In: {ICML}. {ACM} International Conference
  Proceeding Series, vol.~119, pp. 713--719 (2005)

\bibitem{sarwar2001item}
Sarwar, B.M., Karypis, G., Konstan, J.A., Riedl, J.: Item-based collaborative
  filtering recommendation algorithms. In: {WWW}. pp. 285--295 (2001)

\bibitem{SeoHYL17}
Seo, S., Huang, J., Yang, H., Liu, Y.: Interpretable convolutional neural
  networks with dual local and global attention for review rating prediction.
  In: RecSys. pp. 297--305 (2017)

\bibitem{seo2017representation}
Seo, S., Huang, J., Yang, H., Liu, Y.: Representation learning of users and
  items for review rating prediction using attention-based convolutional neural
  network. In: 3rd International Workshop on Machine Learning Methods for
  Recommender Systems (MLRec) (2017)

\bibitem{wang2015collaborative}
Wang, H., Wang, N., Yeung, D.: Collaborative deep learning for recommender
  systems. In: {KDD}. pp. 1235--1244 (2015)

\bibitem{wu2016explaining}
Wu, C., Beutel, A., Ahmed, A., Smola, A.J.: Explaining reviews and ratings with
  {PACO:} poisson additive co-clustering. In: {WWW} (Companion Volume). pp.
  127--128 (2016)

\bibitem{zhang2014explicit}
Zhang, Y., Lai, G., Zhang, M., Zhang, Y., Liu, Y., Ma, S.: Explicit factor
  models for explainable recommendation based on phrase-level sentiment
  analysis. In: {SIGIR}. pp. 83--92 (2014)

\bibitem{zheng2017joint}
Zheng, L., Noroozi, V., Yu, P.S.: Joint deep modeling of users and items using
  reviews for recommendation. In: {WSDM}. pp. 425--434 (2017)

\end{thebibliography}
%


%
%
%
%
%
\end{document}